\documentclass[12pt, A4paper]{article}
\usepackage[a4paper, total={6in, 8in}]{geometry}
\usepackage{graphicx}
\usepackage{dcolumn}
\usepackage{bm}
\usepackage{rotating}
\usepackage{longtable}
\usepackage{float}
\usepackage{eucal}
\usepackage{csquotes}
\usepackage{xcolor}
\usepackage{amsmath}
\usepackage{amssymb}
\usepackage[numbers]{natbib}
\usepackage[pagewise]{lineno}
\usepackage{amstext}
\usepackage{authblk}

\title{Existance of octupole correlation in \textsuperscript{116}Sn}

\author[1]{\small Prithwijita Ray\footnote{email: prithwijitaray@gmail.com}}
\author[2]{\small H. Pai}
\author[3]{\small S. Chakraborty}
\author[4]{\small A. Mukherjee}
\author[5]{\small S. Rajbanshi}
\author[6]{\small Sajad Ali}
\author[7]{\small G. Gangopadhyay}
\author[3]{\small S. Bhattacharyya}
\author[3]{\small G. Mukherjee}
\author[3]{\small C. Bhattacharya}
\author[3]{\small Soumik Bhattacharya}
\author[9]{\small R. Banik}
\author[10]{\small S. Nandi}
\author[11]{\small R. Raut}
\author[11]{\small S. S. Ghugre}
\author[12]{\small S. Samanta}
\author[11]{\small S. Das}
\author[11]{\small S. Chatterjee}
\author[4]{\small A. Goswami\footnote{Deceased}}

\affil[1]{\small Department of Physics, ABN Seal College, Coochbehar-736101, India}
\affil[2]{\small ELI-NP, Horia Hulubei National Institute for R$\&$D in Physics and Nuclear Engineering, Bucharest-Magurele, 077125, Romania}
\affil[3]{\small Variable Energy Cyclotron Centre, HBNI, 1/AF Bidhan Nagar, Kolkata-700064, India }
\affil[4]{\small Saha Institute of Nuclear Physics, HBNI, 1/AF Bidhan Nagar, Kolkata-700064, India}
\affil[5]{\small Department of Physics, Presidency University, Kolkata-700073, India}
\affil[6]{\small Department of Physics, Government General Degree College at Pedong, Kalimpong-734311, India}
\affil[7]{\small Department of Physics, University of Calcutta, Kolkata-700009, India}
\affil[9]{\small Department of Physics, Institute of Engineering \& Management, Kolkata-700091, India}
\affil[10]{\small Argonne National Laboratory, Illinois, USA}
\affil[11]{\small UGC-DAE Consortium for Scientific Research, Kolkata centre, Kolkata 700098, India}
\affil[12]{\small Hebrew University of Jerusalem, Jerusalem, Israel}

\date{\small \today}
\begin{document}
\maketitle
\vspace{-1cm}
\begin{abstract}
The negative parity states in $^{116}$Sn have been investigated in terms of octupole correlation. The same is probed by using the Indian National Gamma Array (INGA) facility at Variable Energy Cyclotron Centre, Kolkata using the reaction, $^{114}$Cd($\alpha$, 2n)$^{116}$Sn at 34 MeV energy. Three new $\gamma$-transitions relevant to the present investigation are reported and the spin-parities of the associated levels are assigned based on the DCO-ratios and polarisation measurements. The interband transitions between the positive parity ground band and the negative parity band are newly observed and the corresponding extracted ratio of transition probability, B(E1)/B(E2) indicated the existence of octupole correlation in these nuclei. The enhanced transition rates for E1 and E3 transitions between these opposite parity bands and corroboration of soft octupole deformation ultimately aid the onset of octupole excitation in this isotope.

\end{abstract}

\section{Introduction}

Singly magic even-mass Sn isotopes near A$\sim$120 have shown a conjuncture of a variety of excitation modes.
Among them, the following are commonly observed: the trend of a ground state band comprising of few levels occurring due to phonon multiplets, positive parity rotational bands due to proton 2p-2h excitation above the shell closure and negative parity levels stated to occur for neutron quasiparticle excitation \cite{bron,poel,chat}. In $^{116}$Sn, the negative parity states have been said to have arisen due to two quasi-neutron excitation based on the $g^{-1}_{7/2}h_{11/2}$ and $d^{-1}_{5/2}h_{11/2}$ configuration \cite{poel, chat}. However, the strong E1 and E3 transitions from these negative parity states as seen in $^{116}$Sn may also indicate the existence of an octupole excitation. This phenomena generate reflection-asymmetric pear-like shapes due to the interaction between the orbitals of opposite parity and having angular momentum differing by 3$\hbar$ near the Fermi surface and arise when the number of nucleons is equal to 34, 56, 88 and 134, the so-called octupole magic numbers \cite{testov}. In both the static octupole deformation or octupole correlation, permanent electric dipole moments are produced which again shows variation with mass. Transition probability, in this context, plays a crucial role in determining if the octupole-octupole interaction is strong enough to produce stable deformation. An observable signature of octupole excitation in even-even deformed nuclei is the presence of a rota­tional band of interleaved positive and negative parity levels. On the other hand, a parity doublet is a signature of octupole deformation in odd-mass nuclei and is defined as a pair of almost degenerate states with the same spin, opposite par­ities, and a large connecting E3 matrix element.
\par
The existence of the enhanced electric dipole and highly collective octupole transition rates are the primary and direct proof for such correlations \cite{schak, subh}. 
Also, the measurement of dipole and quadrupole moment differentiates between the octupole vibration and octupole correlation where lowering of the negative parity band results in the case of the latter \cite{dae2022}.\par
Quite a handful of both theoretical and experimental works have also been done concerning the octupole correlations in the A$\sim$140 regime for Ba, Cs and Xe nuclei \cite{ba,cs, xe}. However, for the A$\sim$120 region, Cottle et. al have predicted an upper mass bound for observation of octupole bands which is at N=66, N=70/72 and N=74 for the lighter Te, Xe and Ba isotopes respectively~\cite{cott} based on the systematics on 5$^-$ and 7$^-$states. Another notable work, in this regard, on $^{124,125}$Ba identifies the connecting E1 transitions to have been raised due to the inclusion of octupole correlations with the negative parity states arising due to $(d_{5/2}+g_{7/2}),h_{11/2}$ interpretation \cite{maso}. This type of mixing enhances the E1 transition rates which are seen in lighter Xe isotopes in the same mass region~\cite{liu}. All these serve as the motivation for testing the negative parity band in $^{116}$Sn for the octuple correlation.

\section{Experiment and data analysis}
The fusion evaporation reaction $^{114}$Cd($\alpha$,2n)$^{116}$Sn was carried out in the Indin National Gamma Array (INGA) campaign at Variable Energy Cyclotron Center, Kolkata. 34-MeV $\alpha$ ion-beam supplied from K-130 cyclotron was made to impinge upon a 6.22 mg/cm$^2$ thick self-supporting target. The $\gamma$-rays emitted by the fusion evaporation reaction mechanism were captured by seven compton-suppressed clover detectors which were positioned at three different angles(40$^\circ$, 90$^\circ$, 125$^\circ$). The time-stamped list-mode data was recorded by Digital Data AQuisition (DDAQ) system consisting of a pixie-16 digitizer-based module. Each pixie-16(having 16 channels) module, manufactured by XIA LLC, can host signals from three clover detectors (3$\times$ 4=12 signals from crystals and 3 for each ACS signal). 
A set of codes, IUCPIX, developed at UGC-DAE Consortium, Kolkata was used to reduce the list-mode data into a single data stream~\cite{sdas,rr2}. Further analysis was done using RADWARE and INGASORT packages~\cite{radware, inganew}. For the spin-parity assignment of the decaying levels DCO-ratio and polarization asymmetry measurements were used.\par
For The DCO-ratio calculation, an asymmetric matrix was constructed using the counts registered in the detectors at 125$^{\circ}$ and 90$^{\circ}$ with the beam axis. For the experimental calculation of DCO-ratio, the following relation is used:
\begin{equation}
\label{rdco}
R_{DCO} = \frac{I_{\gamma_1} ~ at~ 125^{\circ}, ~gated ~by ~\gamma_2 ~at ~90^{\circ}}
               {I_{\gamma_1} ~at ~90^{\circ}, ~gated ~by ~\gamma_2 ~at ~125^{\circ}}
\end{equation}
Where I$_{\gamma_1}$ and I$_{\gamma_2}$ are relative intensities of two coincident $\gamma$-transitions. This ratio is validated for known transitions and is compared with theoretical values using ANGCOR code~\cite{angcor}. The width of the sub-state population (${\sigma}/{J}$)of the present experiment as required for the theoretical R$_{DCO}$ ratio is found to be 0.37. For the present detector arrangement, the DCO-ratio for a pure dipole (quadrupole) transition gated by a stretched quadrupole (dipole) transition is 0.70 (1.50). This is in agreement with measured values as obtained in $^{114}$Te produced in another reaction but having the same set-up. For pure quadrupole transition (774.8-keV, 4$^{+}$ $\rightarrow$ 2$^{-}$) and dipole transition (936.2-keV, 7$^{-}$ $\rightarrow$ 6$^{+}$) the raios are 1.60(0.18) and 0.70(0.08), respectively~\cite{ray2}.
For the determination of polarization asymmetry($\Delta_{\text{asym}}$) two asymmetry matrices were required where coincident $\gamma$-transitions detected at 90$^{\circ}$ were on one axis and transitions detected by all others were on the other axis. $ \Delta_{\text{asym}} $ is expressed as
\begin{equation}
\label{ipdco}
\Delta_{\text{asym}} = \frac{a(E_\gamma) N_\perp - N_\parallel}{a(E_\gamma)N_\perp + N_\parallel},
\end{equation}
where $N_\perp$ and $N_\parallel$ are the number of Compton scattered $\gamma$-rays in parallel and perpendicular directions with respect to the reaction plane (consisting of the beam and the primary incident gamma-ray) and a(E$_{\gamma}$) is called the geometrical correction factor and is given by 
\begin{center}
$a(E_\gamma) = \cfrac{N_{\|}}{N_{\perp}}$
\end{center}
A positive value of polarization asymmetry indicates an electric ($E$) type transition whereas, a negative value indicates a magnetic ($M$) type transition and zero for the mixed type of transition.

\begin{figure}
\begin{center}
\includegraphics[trim=0cm 0cm 0cm 0cm, width=1\columnwidth, angle = 0]{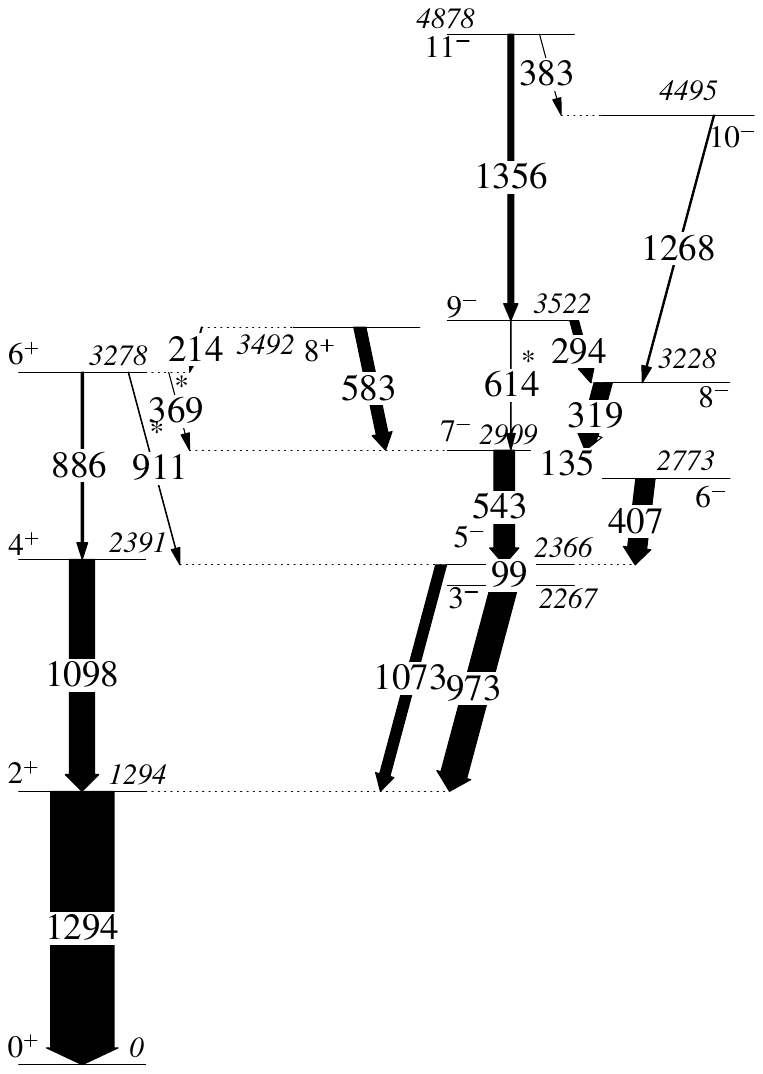}
\caption{The partial level scheme relevant to the present octupole correlation in $^{116}$Sn is presented. The width of the arrows is in accordance with the intensities of the associated transitions. The newly observed $\gamma$-transitions are marked with asterisks(*).}
\label{oct}
\end{center}
\end{figure}
\section{Experimental Results}

\begin{figure}
\begin{center}
\includegraphics*[width=13.0cm, angle = 0]{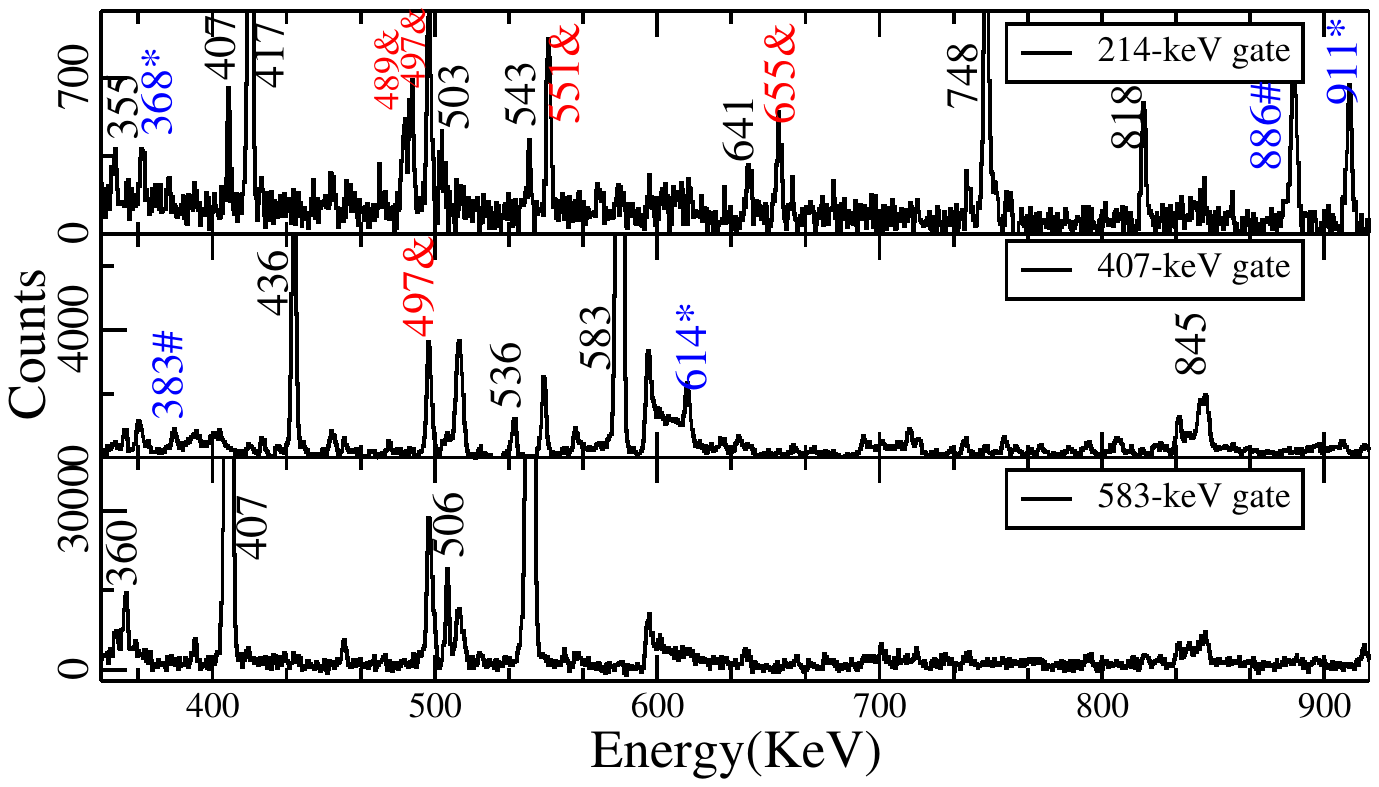}
\vspace{-1cm}
\caption{Gated coincidence spectra showing relevant $\gamma$-transitions. Blue coloured $\gamma$-transitions are marked with asterisks(*) are newly observed whereas hashtag marked ($\#$) are newly observed in this reaction but are previuosly reported (\cite{raythesis}). Black-coloured unmarked ones are previously reported in $^{116}$Sn. The red ampersand marked($\&$) transitions are lines of $^{115}$Sn which were produced in another channels in the fusion evaporation reaction.}
\label{dae_mod2}
\end{center}
\end{figure}

The relevant negative parity band in $^{116}$Sn nuclei is revisited and modified with the placement of new interband and intraband transitions. 
The previously established level scheme of $^{116}$Sn as shown in Ref.~\cite{ray2, raythesis} is updated and presented in Fig.~\ref{oct} relevant to the present analysis. The $\gamma$-$\gamma$ coincidence spectra are shown in Fig.~\ref{dae_mod2} in support of the placement of relevant $\gamma$-transitions.

The level scheme has been constructed using coincidence analysis and intensities of respective $\gamma$-transitions are normalized with respect to the 1293.8-keV transition, with I$_\gamma$ = 100.0. The spin-parities are acquired using DCO-ratios, and polarization asymmetry measurements. The details of the newly observed transitions are discussed below and the rest can be found in Ref.~\cite{ray2,raythesis}.

\subsection{613.5-keV $\gamma$-transition}
\begin{figure}
\begin{center}
\includegraphics*[width=10.0cm, angle = 0]{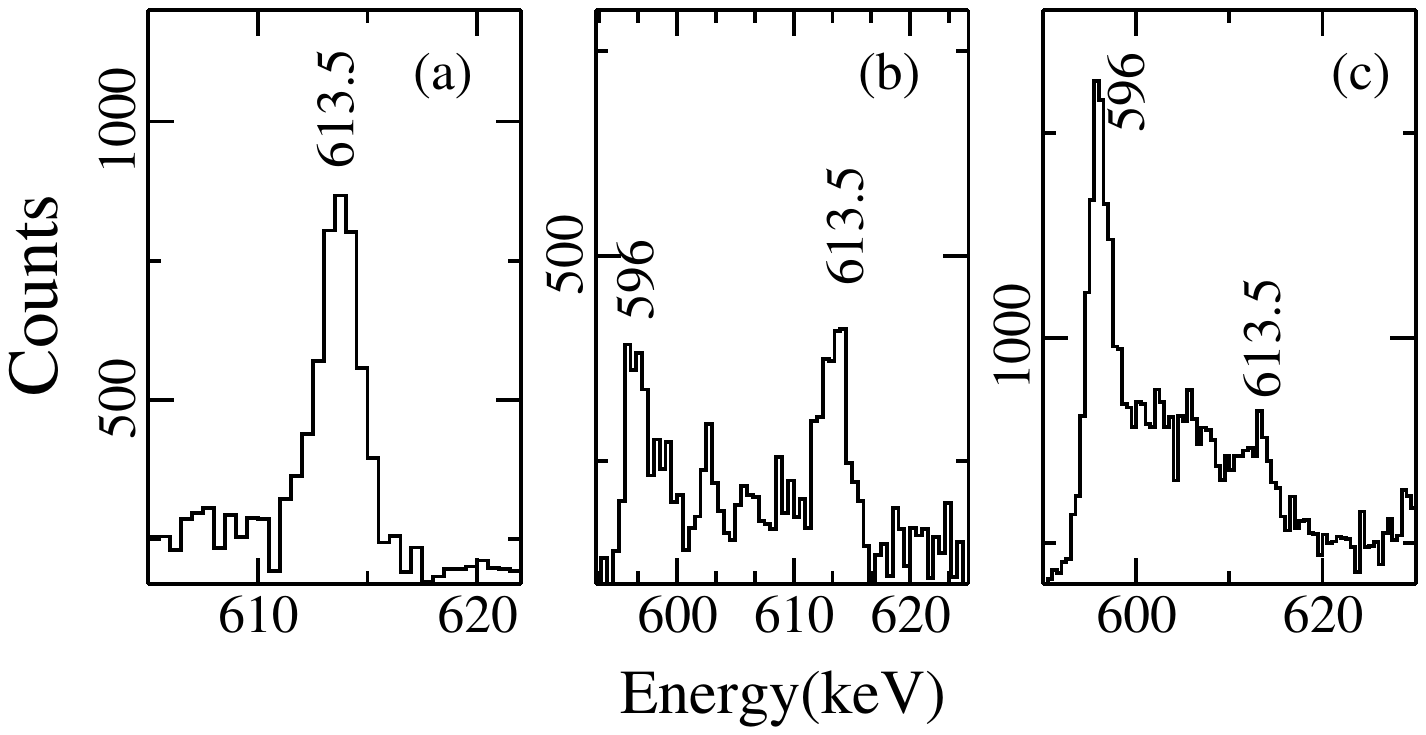}
\caption{Coincidence spectra showing the existence of 613.5-keV decaying from the state at 3521.9-keV. The gated spectra are as follows:(a) 1356.1-keV, (b) 235.1-keV, (c) 844.6-keV. Note that in sub-figures (b) and (c), the presence of 596-keV line is arising due to inelastic neutron scattering from clover detectors. Due to the subsequent energy loss of the neutron, the non-symmetric nature is observed.}
\label{613}
\end{center}
\end{figure}

There exists another 613-keV in $^{115}$Sn nuclei \cite{heavyion} and a closer counterpart in $^{116}$Sn decaying from a state at energy 5495-keV and spin-parity 13$^+$ (not shown in Fig.~\ref{oct}, but can be found in Ref \cite{raythesis}). The existence of this transition decaying from the 3521.9-keV state is depicted in Fig.~\ref{613}. The relative intensity and DCO-ratio as calculated for this transition are 0.83(6) and 0.92(11) respectively.

\subsection{368.8-keV $\gamma$-transition}
\begin{figure}
\begin{center}
\includegraphics*[width=13.0cm, angle = 0]{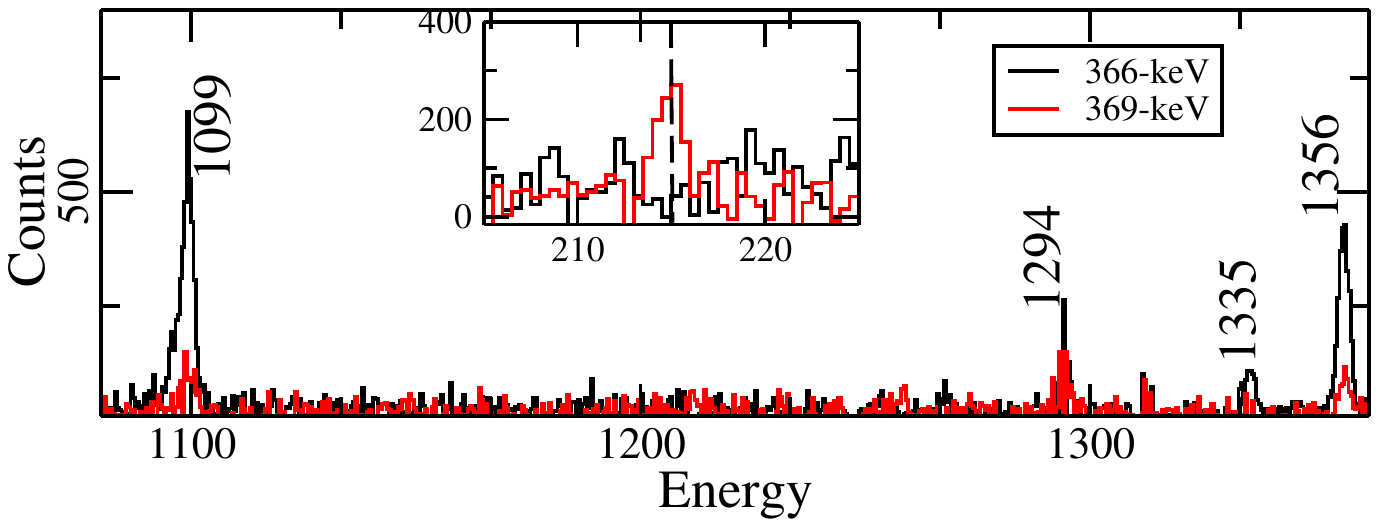}
\caption{Coincidence spectra distinguishing two closely spaced lines 366.2- and 368.8-keV. The inset shows the 214.6-keV transition in 368.8-keV gated spectra (not in 366.2-keV) in support of the present level scheme.}
\label{dae}
\end{center}
\end{figure}

This newly observed dipole transition connecting the negative parity band to the positive parity ground band level is of importance regarding the octupole correlation. However, there exists another transition of energy 366.2-keV decaying from a higher lying 15$^-$ state \cite{raythesis, heavyion}. The existence of this newly found transition is explicitly shown in the coincidence gated spectrum in Fig.~\ref{dae}. The observation of this state is characteristic of the population of numerous non-yrast states in this reaction where an $\alpha$-beam projectile was used. 
The relative intensity of this transition as measured by taking 1293.8-keV as 100 is 0.11(2). The DCO-ratio in the quadrupole gate is found to be 0.67(16)confirming the dipole nature of this transition.

\subsection{911.4-keV $\gamma$-transition}
\begin{figure}
\begin{center}
\includegraphics*[width=13.0cm, angle = 0]{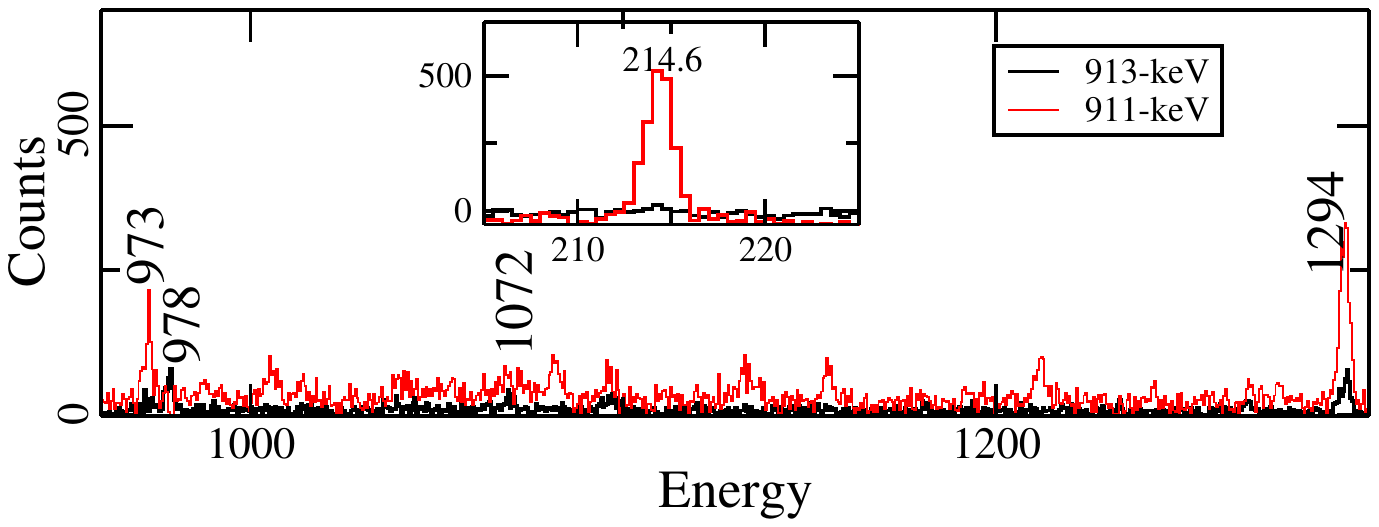}
\caption{Gated coincidence spectra distinguishing two closely spaced lines 911- and 913-keV. The inset shows the 214.6-keV transition in 911-keV gated spectra (not in 913-keV) in support of the present level scheme.}
\label{911}
\end{center}

\begin{center}
\includegraphics*[width=13.0cm, angle = 0]{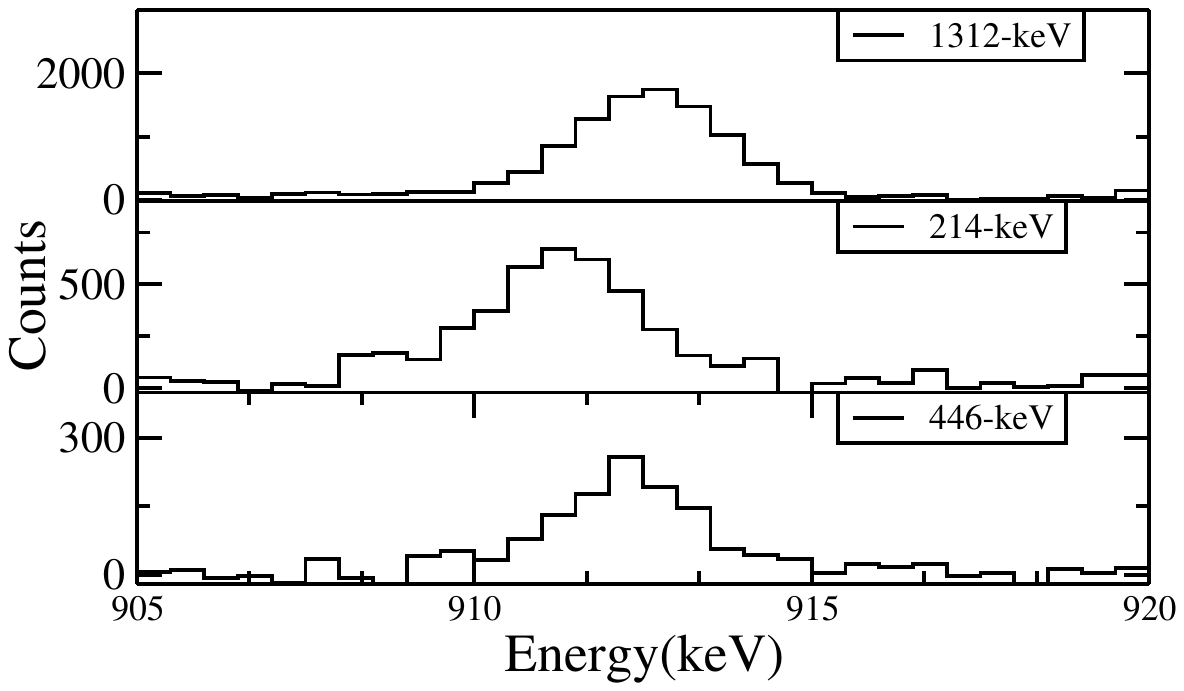}
\caption{Various gated spectra are shown in support of 911-keV transition in the present level scheme.}
\label{diff}
\end{center}
\end{figure}

This connecting E1 transition is also newly observed with relative intensity and DCO-ratio as 0.60(9) and 0.74(30), respectively. The presence of this state separated from two other closely spaced lines in $^{115}$Sn and $^{116}$Sn are shown in Fig.~\ref{911} and \ref{diff}.
\section{Discussion}
\begin{figure}
\hspace{2cm}\begin{center}
\includegraphics[width=80mm]{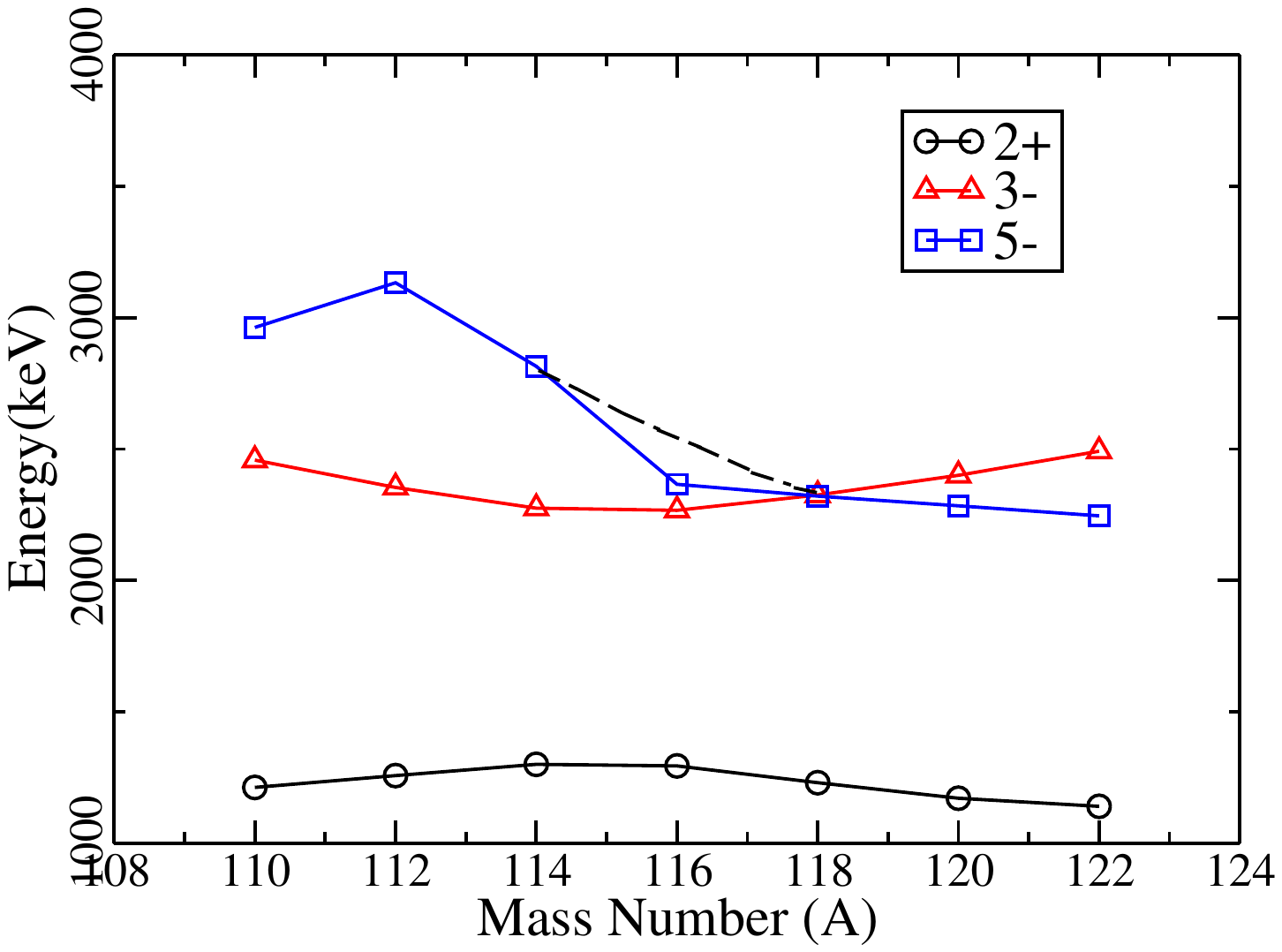}
\caption{\label{systematics} The systematics of 3$^-$ and 5$^-$ energy states in even-even Sn isotopes. Data are taken from NNDC \cite{nndc}.}
\end{center}\end{figure}

The negative parity states in neighbouring nuclei are reported to arise due to quasi-particle excitation. However, the presence of suitable intruder and normal parity subshell is suggestive of octupole correlation as well \cite{maso,piep}. The systematics of energy values of the 3$^-$ and 5$^-$ states in even-even $^{110-122}$Sn isotopes are primarily shown in Fig.~\ref{systematics}. The sudden dip in the energy value in $^{116}$Sn instead of a smooth variation accounts for extra binding energy for the same. This lower-than-expected energy value could be due to significant octupole interaction which could compete with particle alignment very similar to the case of $^{144}$Ba \cite{ba}.

\begin{figure}
\begin{center}
\includegraphics[width=80mm]{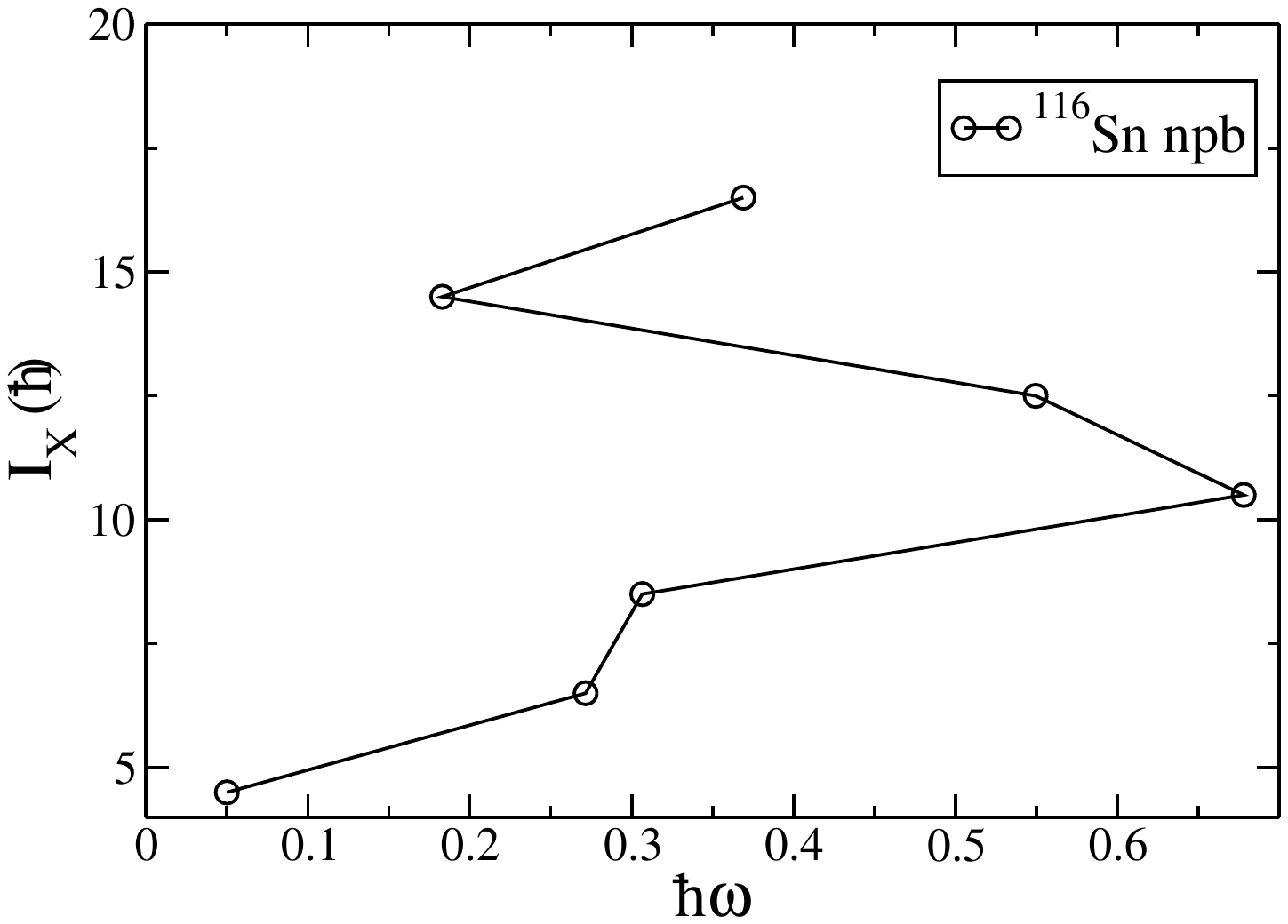}
\caption{\label{alignment} The total aligned angular momentum for the negative parity band in $^{116}$Sn.}
\end{center}
\end{figure}

To have a better understanding of such structural aspects for the negative parity band in the $^{116}$Sn isotope, the alignment plot is shown in Fig.~\ref{alignment}. 
The x-component of the total angular momentum on the rotation axis (I$_x$) is given as
\begin{equation}
I_x=\sqrt{(\frac{I_i+I_f}{2}+\frac{1}{2})^2-K^2}
\end{equation}
Where, K, the projection of angular momentum on the symmetry axis is taken as 0. The $\hbar \omega$ is evaluated as E$_\gamma$/2, for $\Delta$I=2 in-band transitions~\cite{beng,wisn}.\par
A gradual increase of alignment near 0.3MeV is seen similar to $^{150}$Sm expected for octupole-deformed nuclei \cite{urban87}. As we move over 0.3MeV the alignment also increases and at sufficiently higher frequency particle excitation overpowers the octupole collectivity and a back bending is observed at a frequency of 0.67MeV. 
The large (10$\hbar$) gain in alignment could be due to shape changes related to the rotational alignment of nucleon pairs at higher energy regions similar to $^{146}$Ba~\cite{ba}. The question on octupole correlation in the lower and mid-energy remains could still have significance. 

The enhanced transition rate of the E1 and E3 transitions is required for octupole correlation. However, due to the limitation of lifetime measurement, an indirect approach is taken. The ratio of transition rates, [B(E1)/B(E2)], for the associated electric dipole(E1) and electric quadrupole(E2) transition and dipole moment, are computed as per the following equations. 
\begin{equation}
\frac{B(E1)}{B(E2)} = \frac{1}{1.3\times 10^6} \frac{E_\gamma ^5(E2)}{E_\gamma ^3(E1)} \frac{I_\gamma(E1)}{I_\gamma(E2)} [fm^{-2}],
\end{equation}

\begin{equation}
D_0 = Q_0 {\left[ \frac{5}{16} \frac{B(E1)}{B(E2)} \right]}^\frac {1}{2} [efm],
\end{equation}
Where the energies are in MeV and $Q_0$ is the associated quadrupole moment \cite{butl, schak}. For $^{116}$Sn, the qudrupole moment ($Q_0$) is taken as 1.45(21)b from Ref.~\cite{rama}. 
Using the above, the ratio of transition probability and intrinsic electric dipole moment as calculated for this nucleus is tabulated (Table~\ref{tab}) and compared with other nuclei in the same mass region.

\begin{table}[!t]
\centering
\caption{\label{tab} Calculated level energies (E$_i$), spin-parities (J$_i^\pi$, J$_f^\pi$), ratio of transition rates (B(E1)/B(E2)), electric dipole moments (D$_0$) for the present negative parity band and the reported octupole bands in other nuclei near A$\sim$120 mass region.}
\begin{tabular}{ccccc}
\hline\hline
Nucleus & \text{$E_i$} (keV)&
\textbf{$J_{i}$$^{\pi}(\rightarrow J_i^\pi$)} & 
$\frac{B(E1)}{B(E2)}(\times 10^{-6} fm^{-2})$ & D$_0(efm)$ \\
\hline
$^{116}$Sn &	3278&	6$^+$($\rightarrow 5^-$)&	0.09(2)&0.02(4)\\
 &	&	6$^+$($\rightarrow 7^-$)&	0.24(4)&0.04(7)\\
$^{108}$Te \cite{lane} &	2997&	7$^-$&	0.04(5)&	0.02(13)\\
&3661	&	9$^-$&	0.19(13)	&0.05(17)\\
&4491	&	11$^-$&	0.34(3)&	0.06(3)\\
$^{116}$Xe \cite{degr} &	2982&	9$^-$&	0.04	&\\
$^{117}$Xe \cite{liu} &	989&	13/2$^+$&	0.02	&0.02\\
&	1553&	17/2$^+$&	0.03	&0.02\\
&	2178&	21/2$^+$&	0.04	&0.03\\
&	2835&	25/2$^+$&	0.06	&0.04\\
&	3502&	29/2$^+$&	0.11	&0.05\\
$^{124}$Ba \cite{maso} &	2262&	7$^-$&	0.04(1)	&\\
	&	2722&	9$^-$&	0.04(2)	&\\
$^{125}$Ba \cite{maso} &	931&	13/2$^+$&	0.03(8)	&\\
 &	1253&	15/2$^+$&	0.01(7)	&\\
 &	1970&	19/2$^+$&	0.07(2)	&\\
&	2679&	23/2$^+$&	0.18(3)	&\\

\hline\hline

\end{tabular}	\\
\begin{flushleft}
\begin{flushleft}

\end{flushleft}
\end{flushleft}
\end{table}

Both the B(E1)/B(E2) ratio and dipole moment in $^{116}$Sn value is showing similar values to the Ba and Xe isotopes in the mass region \cite{zhu95, zhu97}. This increased octupole interaction near N$\approx$66 for the Sn nucleus follows the trend as seen in Cottle et. al \cite{cott}. 
To further concrete the evidence for octupole excitation, the octupole deformation parameter ($\beta _3$) is calculated from the B(E3)/B(E2) ratio for the 1073 and 99-keV decaying from the 5$^-$ state and found to be 0.10(6). The same is also evaluated using the average value of the dipole moment[D$_{avg}$=0.03(8)] observed in the present analysis and using the following equation~\cite{frie}:
\begin{equation}
D_{avg} = 0.000687AZ \beta _2 \beta _3 [efm],
\end{equation}
The calculated $\beta _3$ is 0.07(2) consistent with octupole excitation.

\subsection{Other observations}
Evidence of strong octupole correlation in even-even nuclei is the observation of interleaved positive and negative parity states and the clustering of the states is indicative of the same~\cite{casten_AK}. For octupole correlations, the odd-even spin staggering [S(I)=$E(I^-)-\frac{E((I+1)^+)+E((I-1)^+)}{2}$] has minima at odd spins contrary to the quadrupole staggering seen in positive parity $\gamma$-band~\cite{ray}.
\begin{figure}
\hspace{2cm}\begin{center}
\includegraphics[width=100mm]{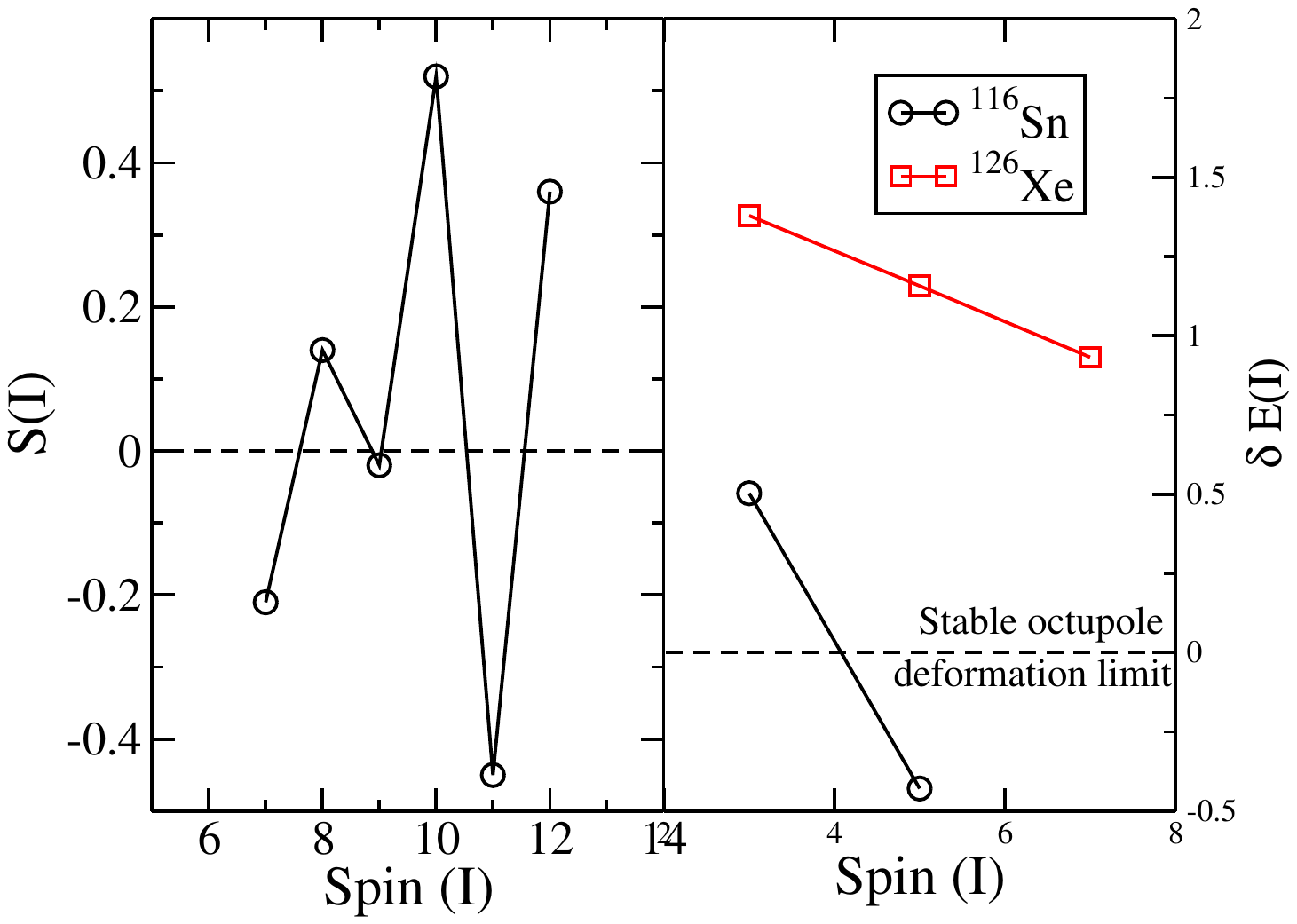}
\caption{\label{S_I} (Left)The odd-even staggering[S(I)] of experimentally obtained negative parity states in $^{116}$Sn. (Right) Plot of the energy displacement $\delta$E(I)in MeV as obtained in $^{116}$Sn and $^{126}$Xe~\cite{schak2}.}
\end{center}\end{figure}
The plot of S(I) vs. spin (I) in Fig.~\ref{S_I} for $^{116}$Sn shows a similar type of clustering as in $^{122, 124, 126}$Ba but only with a smaller magnitude~\cite{casten_AK}. \par
In this context, the energy displacement[$\delta$E(I)=$E(I^-)-\frac{(I+1)E(I-1)^++IE(I+1)^+}{2I+1}$] is also shown for the simplex band (s=+1) in Fig.~\ref{S_I} indicating the presence of reflection asymmetry in $^{116}$Sn.

\section{Summary}
Spectroscopic analysis has been done on $^{116}$Sn nuclei produced in the $\alpha + ^{114}$Cd reaction at 34-MeV using the INGA facility at VECC.  
Three new $\gamma$-transitions associated with the negative parity band are reported along with the existing ones and their electromagnetic nature has been deduced based on the DCO-ratio and polarization measurements. The B(E1)/B(E2) ratio, dipole moment and octupole deformation parameter as calculated for the newly observed connecting transition are suggestive of octupole correlation. The odd-even spin staggering is also indicative of the same. However, more experimental observations of E1 transitions could be explored to gauge such correlations in the vicinity.

\section*{Acknowledgment}
The tremendous support received from all the INGA collaborators especially from VECC and UGC-DAE consortium, Kolkata during the setting up of the array and DDAQ system is highly appreciated. Consistent help from the operating staff of VECC for providing the stable beam throughout the INGA run is acknowledged. INGA is partially funded by the Department of Science and Technology, Government of India (No. IR/S2/PF-03/2003-II). S. R. would like to acknowledge the financial assistance from the University Grants Commission - Minor Research Project [No. PSW-249/15-16 (ERO)]. G.G. acknowledges the support provided by the University Grants Commission, Departmental Research Support (UGC-DRS) program. H.P. is grateful for the support of the Ramanujan Fellowship research grant under SERB-DST (SB/S2/RJN-031/2016).\\


\end{document}